\documentclass[preprint,amsmath,amssymb]{revtex4}

\usepackage{graphicx}
\usepackage{dcolumn}
\usepackage{bm}

\begin{document}


\title{Spin-polarized collision of deuterium and tritium: \\ Relativistic Kinematics}

\author{Thomas B. Bahder}
\author{William C. McCorkle}
\affiliation{Charles M. Bowden Research Facility\\
Weapons Sciences Directorate, Army Aviation and Missile Research, Development, and Engineering Center, 
Redstone Arsenal, Alabama  35898-5000}
\author{William V. Dent, Jr.}
\affiliation{Dent International Research, Inc. \\
3000 Turnberry Drive \\
Hampton Cove, AL 35763}
\email{thomas.bahder@us.army.mil}
%
%

\date{\today}

\begin{abstract}
We investigate the relativistic kinematics of the spin-polarized collision of deuterium incident on tritium, producing  ${}^4$He and a neutron.    Within the context of special relativity, we apply the conservation of four momentum  and the conservation of intrinsic spin, which leads to a system of ten equations. We impose initial conditions such that the deuterium is moving along the x-axis, the tritium is stationary at the origin of coordinates, and the classical spin vector of the deuterium (spin magnitude = 1) is along the $+$z-axis, while the classical spin vector for tritium (spin magnitude = 1/2) is along the $-$z-axis.   We expand the ten conservation equations to second order in velocities and we solve them for the velocity components of the neutron, its unit-spin-orientation vector, and the velocity components of the ${}^4$He nucleus, as a function of the incident deuterium energy.  We find that this analytic solution agrees closely with the numerical solution of the ten (unexpanded) equations.  For a given energy of deuterium,  we find that there are two solutions, each solution having a unique velocity for the emitted neutron and helium nucleus.  The two solutions are related to each other by reflection in the plane perpendicular to the deuterium spin and containing the initial deuterium velocity vector.   
\end{abstract}

\maketitle

\section{Introduction}
  
Sources of neutrons are currently used to excite nuclei in an unknown sample of material in order to determine its elemental composition. The technique is called neutron activation analysis, where neutrons are incident on target nuclei and a unique spectrum of gamma rays are emitted for each element that is present in the sample~\cite{Alfassi1990}.  In this technique, measurements are made of emitted gamma rays that are either emitted almost instantaneously (prompt gamma-ray), or, that are delayed. The prompt gamma-rays that are emitted almost instantaneously come from a compound nucleus that is formed when the neutron is captured by the target nucleus in the sample.  The technique is called prompt gamma-ray neutron activation analysis (PGNAA).  When delayed gamma rays are measured, they are emitted by the decay of a radioactive intermediate state formed by the neutron irradiation.  In this case, the technique is called delayed gamma-ray neutron activation analysis (DGNAA).  In both cases,  analysis of the composite emitted gamma ray spectrum often allows a precise determination of the elemental content of the sample~\cite{Alfassi1990}.  A critical component in this type of detection is the source of neutrons.  Ideally, we would want a narrow beam of neutrons that can be directed at the sample so that the damaging effects of neutron radiation are limited to a small solid angle. A reaction that is commonly used for neutron production is the collision of deuterium with tritium~\cite{Ohlsen1967}, 
\begin{equation}
d + t \rightarrow n + {}^4{He} 
 \label{d_t_reaction}%
\end{equation}
However, with unpolarized deuterium and tritium this reaction generally produces an isotropic distribution of emitted neutrons, as opposed to the narrow beam mentioned above.  If the deuterium and tritium nuclei were spin polarized before collision, then the symmetry of the collision, $\overrightarrow T (\overrightarrow d ,\overrightarrow n )\,{}^4He $, would be reduced, and we would expect that there may be preferential directions for the emitted neutrons.   Quantum mechanical calculations of this collision process are quite complicated~\cite{Ohlsen1967,Kulsrud1982}.   However, it is possible to use a considerably simpler approach to analyze some aspects of the kinematics of the spin-polarized deuterium and spin-polarized tritium collision.   We use  conservation of four-vector momentum and conservation of intrinsic spin tensor, within the context of special relativity, to investigate the kinematics of this collision.  

This reaction in Eq.(\ref{d_t_reaction}) has the advantage that it  has a large cross section, which peaks at 5 b~\cite{Ohlsen1967,Bonar1968} for a deuterium energy of 107 keV.   It is believed that at low-energy the reaction, T(d,n)${}^4He$, proceeds via a short-lived intermediate resonant state of ${}^5He$ with a spin angular momentum $J=3/2$, and that this state quickly decays into a neutron and a ${}^4He$ nucleus.  At low energies, the reaction proceeds via zero orbital angular momentum, $l=0$, so that the total spin $J=3/2$ must arise from intrinsic spin of the reactants~\cite{Kulsrud1982}. Therefore, at low energies, the deuterium nuclear spin $s_d=1$,  and tritium nuclear spin, $s_t=1/2$, must be aligned to produce a total angular momentum $J=3/2$ in the initial state.  The products of the reaction are a neutron, with nuclear spin $s_n=1/2$, and  ${}^4He$ which has nuclear spin $s_h=0$.  The total intrinsic spin of the product state (helium and neutron) is $s=1/2$, so the rest of the angular momentum, the quantity $J=3/2-1/2=1$, must be carried off as orbital angular momentum of the product state.  Alternatively, there is also some small amplitude for the reaction to proceed, with zero orbital angular momentum, when the deuterium and tritium spins are anti-aligned.  In this case, there is zero orbital angular momentum carried off by the products.  We consider this latter case in the work below.  

The formalism we use was initially applied by J. L. Synge to discuss elastic collisions (without transmutation of mass) of particles having intrinsic spin~\cite{Synge1965}.  Here, we use Synge's formalism to analyze the fundamentally inelastic collision,    
$\overrightarrow T (\overrightarrow d ,\overrightarrow n )\,{}^4He $, when the reactants have a definite state of spin, in the sense of special relativity, where spin is represented as a four-tensor. The treatment we give below assumes that the deuterium has a zero impact parameter on the tritium, and therefore this limits our analysis to the case where the spin of deuterium and tritium are anti-aligned, i.e., the spins are pointing in opposite directions. Unfortunately, the condition of zero impact parameter is not easily relaxed because of complications in special relativity associated with the speed of propagation of interactions and the requirement of relativistic invariance~\cite{Synge1965}.   As described above, at low energies, the zero impact parameter (zero orbital angular momentum in initial state) is probably the most important in the collision process since the reaction $\overrightarrow T (\overrightarrow d ,\overrightarrow n )\,{}^4He $ will only occur when the deuterium and tritium are in close spatial proximity, due to the short range nature of the nuclear forces.

Therefore, our goal here is to analyze the dependence of the velocity components (angular dependence)  of the emitted neutron and helium nuclei, on the initial conditions, which are specified by initial velocity  and initial spin orientation for deuterium and tritium.   We apply the conservation of four momentum and conservation of intrinsic spin four-tensor.

\section{Conservation of four momentum}

The total four momentum before the collision is equal to the total four momentum after the collision
\begin{equation}
p_k^{(d)}  + p_k^{(t)}  = p_k^{(n)}  + p_k^{(h)} 
 \label{four_vector_conservation}%
\end{equation}
where the superscripts $d$, $t$, $n$, and $h$ label the four momentum vectors for deuterium, tritium, neutron, and helium, respectively.  The subscript $k=1,2,3,4$ labels the space-time components of the four-vector. We follow the convention used by Synge and take the fourth component to be imaginary, thereby omitting the need to explicitly introduce a metric tensor for space-time.  So the momentum four-vector  has the form 
\begin{equation}
p_k = (p_\alpha,p_4)= m (\frac{\gamma}{c} \, u_\alpha, i \, \gamma)
 \label{four_vector_momentum}
\end{equation}
where $\gamma=(1-(u_\alpha \, u_\alpha)/c^2)^{-1/2}$.  Greek indices take values $\alpha=1,2,3$, and Latin indices always take values $k=1,2,3,4$.
We use the convention that repeated indices are summed over their respective ranges.   Here, $u_\alpha = d \, x_\alpha / d \, t$ are the ordinary three velocity components.   Equation~(\ref{four_vector_conservation}) is a conservation law that is valid in any inertial frame of reference.  For our purposes, we assume that the laboratory frame is an inertial frame.  The conservation of four momentum leads to four equations.

\section{Angular Momentum Tensor}  
In four dimensional space-time, the total angular momentum is an antisymmetric second rank tensor, $H_{mn} = H^{orb}_{mn} + H^{spin}_{mn}$, with six independent components, for $m,n=1,2,3,4$, where $H^{orb}_{mn}$ is the orbital angular momentum and $H^{spin}_{mn}$ is the spin angular momentum~\cite{Synge1965}.  The spin is represented by a four tensor because in four dimensions (space-time) there is no four-vector (of angular momentum)  that is dual to the tensor $H_{mn}$.   

As discussed above, we assume that in the laboratory frame of reference the impact parameter for deuterium on tritium is zero, so there is zero orbital angular momentum in the collision, $H^{orb}_{mn}=0$. In what follows, we drop the superscript and we use $H_{mn}$ to denote the intrinsic spin four-tensor, $H^{spin}_{mn}$.  

For a particle that is not experiencing any forces, the intrinsic spin four-tensor is conserved along the particle's world line~\cite{Synge1965}.   When a collision occurs at a space-time point, the sum of the intrinsic spin four-tensors before the collision must be equal to the sum of these tensors after the collision, since the particles do not experience forces before or after the collision. We assume that the range of forces between particles is negligible, so that we take the collision to occur at a point.  The intrinsic spin four-tensor can be expressed in terms of the particle's four velocity, 
\begin{equation}
\lambda_n =  (\frac{\gamma}{c} \, u_\alpha, i \, \gamma)
 \label{four_vector_velocity}
\end{equation}
and a unit spin four-vector, $s_n$, according to~\cite{Synge1965}
\begin{equation}
H_{kl}= i \, \Omega  \, \epsilon_{klmn} \, \lambda_m \, s_n
\label{spin_four_tensor}
\end{equation}
where $\Omega $ is the magnitude of the intrinsic spin, the four-velocity satisfies $\lambda_n \, \lambda_n = -1$ and the unit spin four-vector satisfies, 
\begin{equation}
s_n s_n = 1.   
\label{unit_four_vector_s}
\end{equation}
The Levi-Civita symbol, $\epsilon_{klmn}$, is antisymmetric with respect to interchange of adjacent indices, it satisfies $\epsilon_{1234}=+1$ and has value -1 for odd permutations of its argument, so that 
\begin{equation}
H_{kl}= 
i \, \Omega \,\, \left(\begin{array}{cccc}
 0 &   (s_4 \lambda_3-s_3 \lambda_4) &   (s_2 \lambda_4-s_4 \lambda_2 ) & ( s_3 \lambda_2-s_2 \lambda_3 )  \\
 (s_3 \lambda_4-s_4 \lambda_3)  & 0 &  ( s_4 \lambda_1-s_1
   \lambda_4 ) & (s_1 \lambda_3-s_3 \lambda_1 ) \\
   (s_4 \lambda_2-s_2 \lambda_4 ) & (s_1 \lambda_4-s_4 
   \lambda_1 ) & 0 & (s_2 \lambda_1-s_1 \lambda_2 ) \\
 (s_2 \lambda_3-s_3 \lambda_2 ) & (s_3 \lambda_1-s_1 
   \lambda_3 ) & (s_1 \lambda_2-s_2 \lambda_1) & 0
\end{array}
\right)
\label{spin_four_tensor2}
\end{equation}
The intrinsic spin
four-tensor  has the following frame-independent invariant, 
\begin{equation}
\frac{1}{2} H_{kl} \, H_{kl} =  \Omega^2. 
\label{H_invariant}
\end{equation}
Furthermore, the intrinsic spin four-tensor is orthogonal to the four velocity
\begin{equation}
H_{mn} \, \lambda_n = 0
\label{spin_velocity_orthogonal}
\end{equation}
by virtue of its construction in Eq.~(\ref{spin_four_tensor}).

The four components of the unit spin four-vector, $s_n$, are related to the four-velocity of the particle, $\lambda_n$, by the orthogonality condition~\cite{Synge1965}
\begin{equation}
 \lambda_n  \, s_n= 0
\label{spin_velocity_orthogonal2}
\end{equation}
Equations~(\ref{spin_velocity_orthogonal}) and (\ref{spin_velocity_orthogonal2}) state that the physical spin vector lies in the three-dimensional hypersurface that is the three-dimensional physical space. Equations~(\ref{spin_four_tensor} )  through (\ref{spin_velocity_orthogonal2}) are tensor equations and hence they are valid in any frame of reference.  

The fourth component of the unit spin four-vector, $s_4$ can be expressed in terms of the other three components by use of the orthogonality condition in Eq.~(\ref{spin_velocity_orthogonal2}), 
\begin{equation}
s_4 = \frac{i}{c}(  s_1 \, u_1   + s_2  \, u_2  + s_3 \, u_3 )  
\label{spin_4}
\end{equation}
Using Eq.~(\ref{spin_4}) to eliminate $s_4$ from Eq.~(\ref{unit_four_vector_s}), we have a relation between the three components of spin, $s_\alpha$, and the three velocity, $u_\beta$, which must be satisfied 
\begin{equation}
s_\alpha \, s_\alpha = 1 + \frac{1}{c^2}(  s_\beta \, u_\beta )^2  
\label{spin_velocity_relation}
\end{equation}
If we define three-component quantities by ${\bf s}=(s_1,s_2,s_3)$ and ${\bf u}=(u_1,u_2,u_3)$, then we can write suggestively that the quantity ${\bf s}$ is normalized according to  
\begin{equation}
{\bf s}^2 = 1 + \frac{1}{c^2}(  {\bf s} \cdot {\bf u} )^2  
\label{spin_velocity_relation2}
\end{equation}
Equation~(\ref{spin_velocity_relation}) or (\ref{spin_velocity_relation2}) shows that the three spatial components ${\bf s}$ of the unit spin four-vector, $s_n$, satisfy a normalization relation that depends on the particle's three velocity components, ${\bf u}$.  When the particle is at rest, ${\bf u}= 0 $, the three components satisfy  ${\bf s} \cdot {\bf s}  = 1$, as expected.   In the low velociy limit, we can interpret the three component quantity, ${\bf S} = \Omega {\bf s}$, as the classical spin vector, where $\Omega$ is the magnitude of the particle's spin.   Of course, the transformation properties of the three components, ${\bf S}$, depend on velocity and hence do not transform as a true three-vector.

\section{Intrinsic Spin Conservation}  

We consider the collision, 
$\overrightarrow T (\overrightarrow d ,\overrightarrow n )\,{}^4He $ where the initial conditions are such that both the deuterium and tritium  spins are anti-aligned. The magnitude of total spin for deuterium is assumed to be $\Omega_d=1$ and the magnitude of spin for the tritium nucleus is taken to be $\Omega_t=1/2$.  The products of this inelastic collision are a ${}^4He$ nucleus, whose total spin is $\Omega_{h}=0$,   and a neutron, whose total spin is $\Omega_n=1/2$.   The conservation of intrinsic spin before and after the collision is given by
\begin{equation}
H^d_{mn} +  H^t_{mn} =  H^{h}_{mn} + H^n_{mn}
\label{spin_conservation}
\end{equation}
where $H^{h}_{mn}=0$ because the spin of He is zero.   Each of the intrinsic spin four-tensors in Eq.~(\ref{spin_conservation}) satisfies Eq.~(\ref{spin_velocity_orthogonal}) by virtue of their construction according to Eq.~(\ref{spin_four_tensor}).  Each spin tensor  $H_{mn}$ is a $4\times4$ antisymmetric matrix. Therefore, Eqs.~(\ref{spin_conservation}) constitute six independent equations that relate the spin orientations to the velocity components, before and after the collision.     Equations~(\ref{spin_conservation}) are tensor equations that are valid in any inertial frame of reference. 

\section{Initial Conditions and Solutions}

We have a total of ten conservation equations, where four of them are the conservation of four momentum given by Eq.~(\ref{four_vector_conservation}) and six of them are the conservation of intrinsic spin given by Eq.~(\ref{spin_conservation}).

In solving the kinematic problem of the deuterium-tritium collision, the magnitude of spin for each particle is known, as described above. The orientation of the spins and velocities of deuterium and tritium are initial conditions that can be selected arbitrarily. We solve for the velocity and spin orientation components of the neutron, and the velocity components of the ${}^4$He, in the laboratory frame of reference. For initial conditions, we take the tritium to be stationary at the origin of coordinates with velocity  $u_{\alpha}^{t}=(0,0,0)$, and we take the deuterium to be moving along the $+x$-axis, with velocity components specified by $u_{\alpha}^{d}=(u,0,0)$, where $u>0$.   We take the deuterium spin direction to be along the $+z$-axis, specified by the spin orientation three-vector as $s_\alpha^{d}=(0,0,1)$ and we take the tritium spin orientation three-vector to be $s_{\alpha}^t=(0,0,-1)$. The deuterium is moving such that ${\bf s} \cdot {\bf u} = 0$, so ${\bf s}^2=1$, and the initial condition $s^d_\alpha=(0,0,1)$ satisfies Eq.(\ref{spin_velocity_relation2}).  (As described above, the magnitude of the spin for every particle is fixed.)  We solve for the following quantities: velocity and spin orientation components of the neutron, $u^{n}_\alpha = (u^n_x,u^n_y,u^n_z)$ and  $s^{n}_\alpha = (s^n_x,s^n_y,s^n_z)$, respectively, and velocity of ${}^4$He, $u^{h}_\alpha = (u^h_x,u^h_y,u^h_z)$.   Note that we use subscripts $x,y,z$ and 1,2,3 interchangeably to denote components.

\begin{figure}
\includegraphics{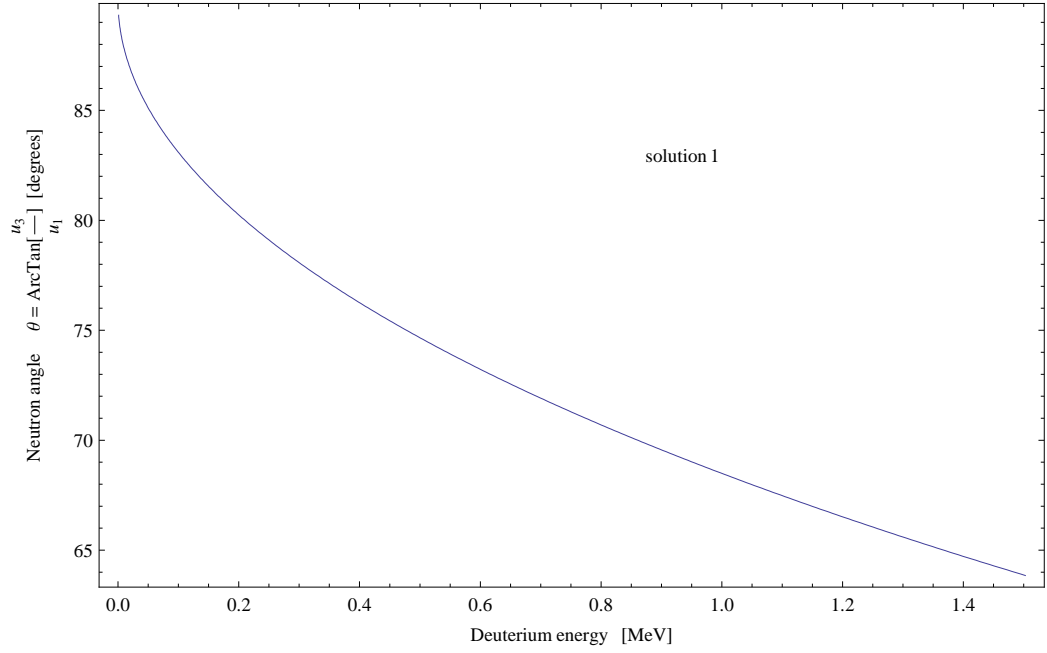}
\caption{The angle of neutron as a function of deuterium kinetic energy in the laboratory frame of reference for solution 1.}%
\label{neutron_angle_soln1}
\end{figure}

\begin{figure}
\includegraphics{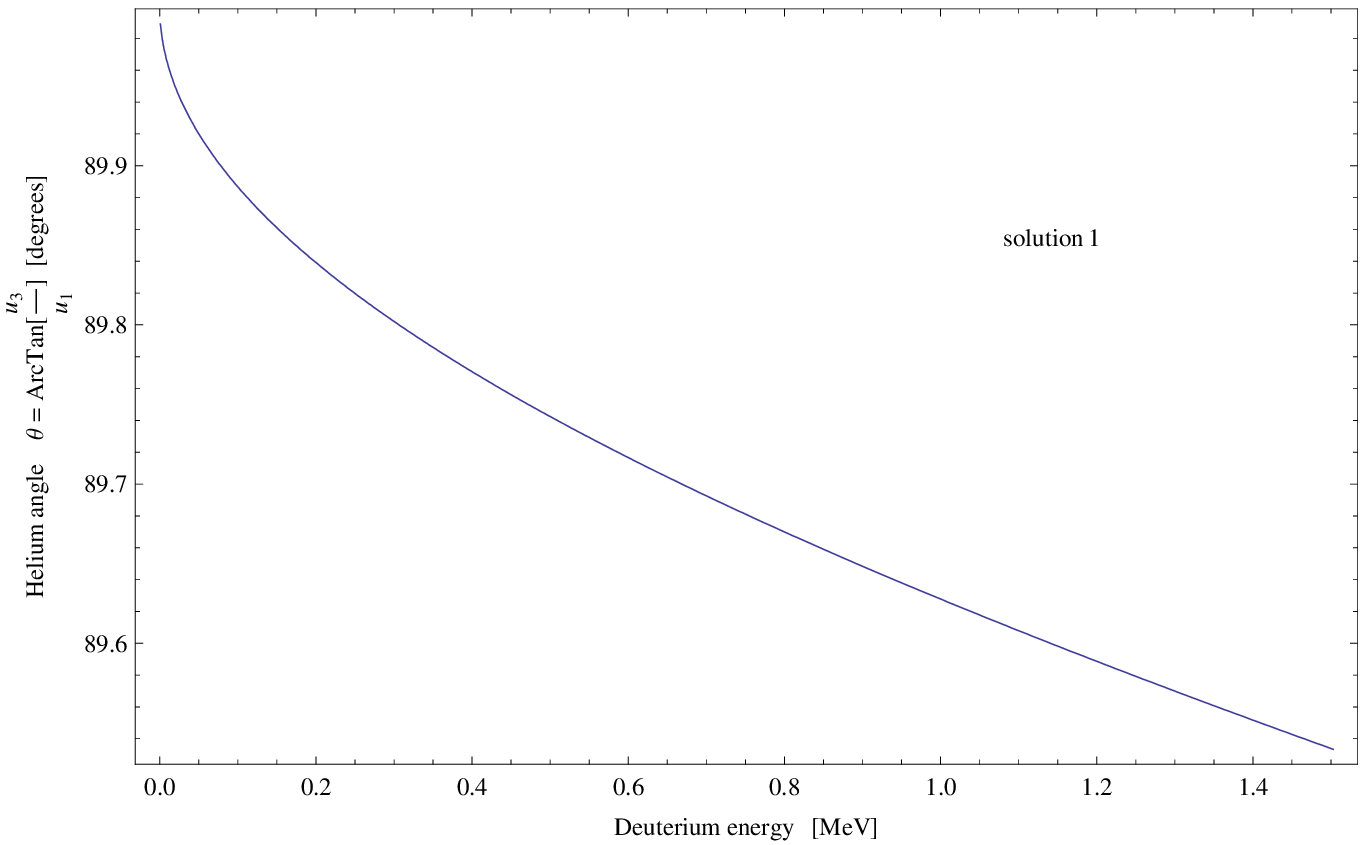}
\caption{The angle of helium as a function of deuterium kinetic energy in the laboratory frame of reference for solution 1.}%
\label{helium_angle_soln1}
\end{figure}

\begin{figure}
\includegraphics{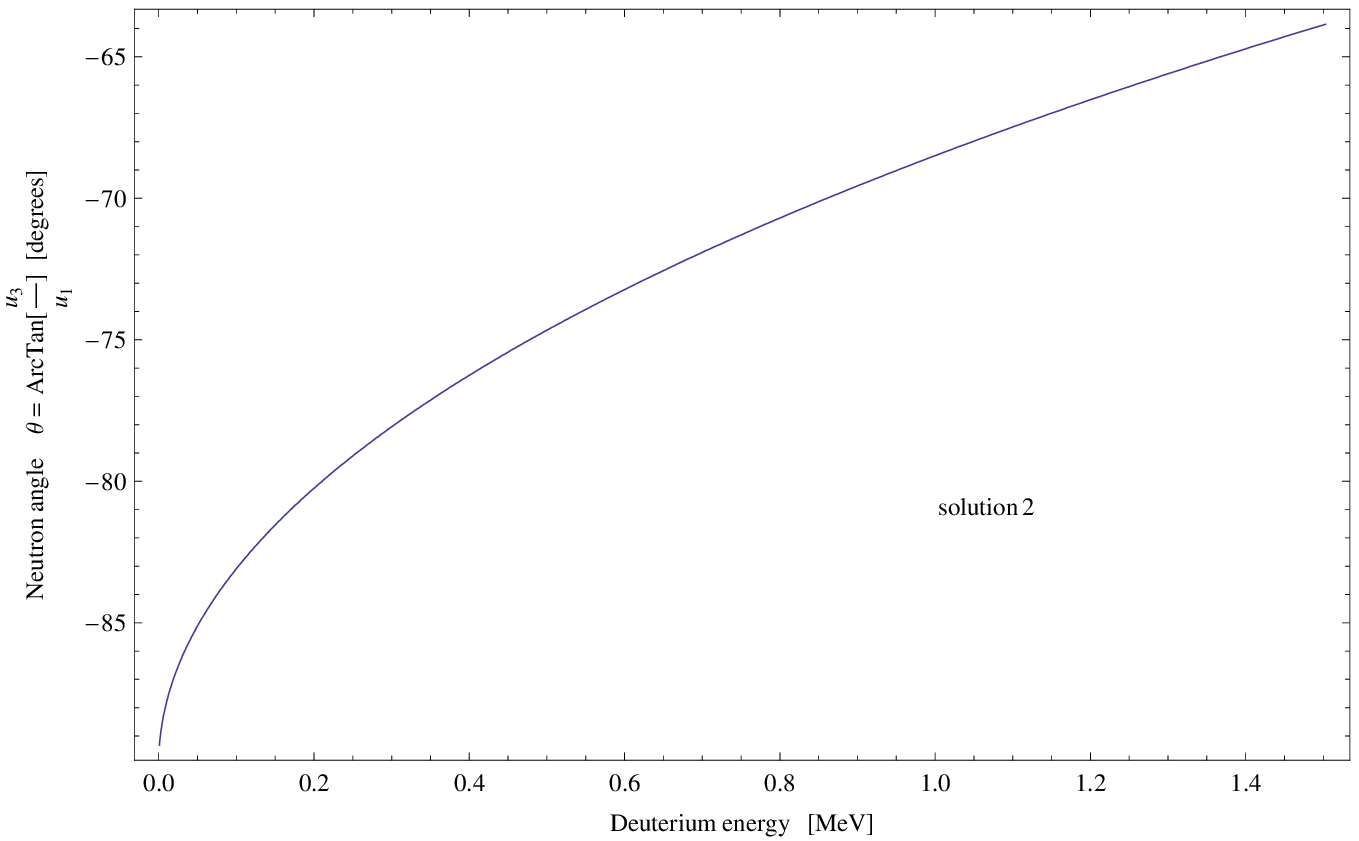}
\caption{The angle of neutron as a function of deuterium kinetic energy in the laboratory frame of reference for solution 2.}%
\label{neutron_angle_soln1}
\end{figure}

\begin{figure}
\includegraphics{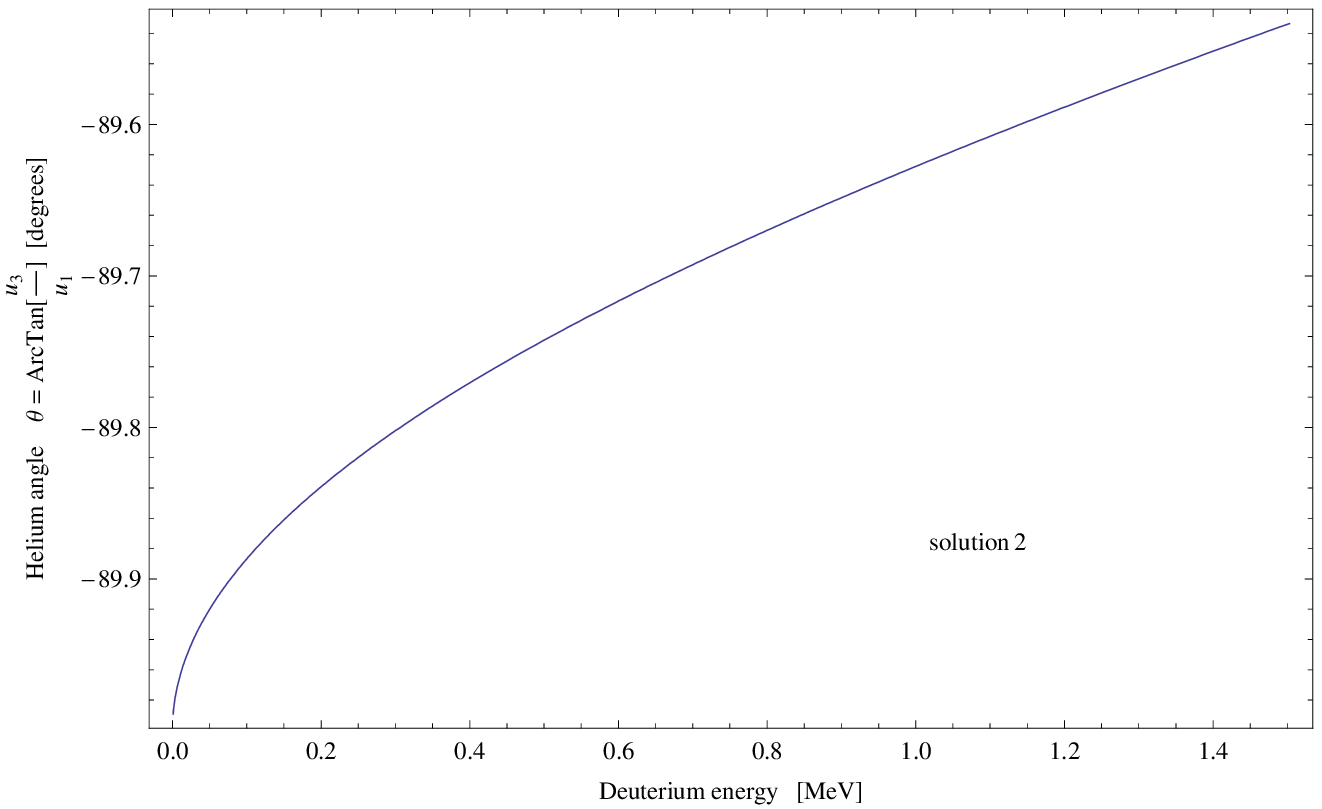}
\caption{The angle of helium as a function of deuterium kinetic energy in the laboratory frame of reference for solution 2.}%
\label{helium_angle_soln1}
\end{figure}

We have a total of ten nonlinear equations, given by Eq.~(\ref{four_vector_conservation}) and (\ref{spin_conservation}).  There are a total of nine unknowns: three velocity components for the neutron, $u^{n}_\alpha$, three velocity components for the ${}^4$He,  and three spin-orientation vector components for the neutron spin, $s^{n}_\alpha$.   A key point is that, before solving the ten Eqs.~(\ref{four_vector_conservation}) and (\ref{spin_conservation}), we must eliminate from Eqs.~(\ref{spin_conservation}) the fourth component of the spin-orientation vector for each particle using Eq.~(\ref{spin_4}).  In this way, only the three spatial components, $s^n_\alpha$, of the spin-orientation four-vector $s^n_m$, appear in the intrinsic spin tensors in Eqs.~(\ref{spin_conservation}).  In the solutions of our ten equations, the spin-orientation vector for the neutron, ${\bf s}^{(n)} = (s^n_x,s^n_x,s^n_x)$,  and the neutron velocity, $u^{n}_\alpha$, must still satisfy  Eq.~(\ref{spin_velocity_relation}) or (\ref{spin_velocity_relation2}). We have checked that our solutions satisfy Eq.~(\ref{spin_velocity_relation}).

Having eliminated the  components, $s^d_4$, $s^t_4$, $s^n_4$, and $s^h_4$, the ten Eqs.~(\ref{four_vector_conservation}) and (\ref{spin_conservation}) are solved by expanding them in a Taylor series to second order in the velocity components, $u^{n}_\alpha$,  $u^{h}_\alpha$ and $u$, which we assume are small quantities compared to the speed of light, $c$.  At least second order in the velocities must be kept to allow for the transmutation of mass, i.e., inelastic collision process.   We find that there are two solutions to these ten equations, and each solution has a unique velocity for the neutron and for the ${}^4$He nucleus. The two solutions, which we call solution 1 and solution 2, are related by reflection in the  $x-y$ plane, which is perpendicular to the deuterium spin, $s_\alpha^{d}=(0,0,1)$,  and  contains the deuterium velocity vector, $u^{d}=(u,0,0)$.  Note that the signs of the velocity components for the two solutions are as follows:  
\begin{eqnarray}
{\rm solution \,\, 1:} \,\,\,\,  &   u^{n}_x > 0, \,\, u^{n}_y = 0,  \,\, u^{n}_z > 0;  \,\,\,\, u^{h}_x < 0, \,\, u^{h}_y = 0, \,\,  u^{h}_z < 0 \nonumber \\
{\rm solution \,\, 2:} \,\,\,\, &   u^{n}_x > 0, \,\, u^{n}_y = 0,  \,\, u^{n}_z < 0;  \,\,\,\, u^{h}_x < 0, \,\, u^{h}_y = 0, \,\,  u^{h}_z > 0 \nonumber 
\end{eqnarray}

Furthermore, we solve numerically the ten nonlinear (unexpanded) equations, given by Eq.~(\ref{four_vector_conservation}) and (\ref{spin_conservation}), and compare these numerical solutions to the analytic solutions.   The two solutions agree very well.

The solutions show that the emitted neutron and helium have velocity vectors that lie in the x-z plane. Figures 1 and 2 show the angle of the emitted neutron and helium, respectively, as a function of incident deuterium kinetic energy, for solution 1.  Figures 3 and 4 show that same quantities for solution 2, which is related to solution 1 by reflection, as described above. The numerical data used in our calculations is given in Table I.

At the maximum cross section, where the deuteron kinetic energy is 107 keV, the angle of the neutron is $\theta_n = \arctan(u_3 / u_1) = 82.85$ degrees and the angle of the helium is $\theta_h = \arctan(u_3 / u_1) = 89.88$ degrees.

\section{Discussion}
Within the context of special relativity, we have applied the conservation of four momentum and the conservation of intrinsic spin to a collision of deuterium with a stationary tritium atom, in the laboratory frame of reference. We have assumed that the deuterium and tritium nuclei have spins that are anti-aligned and have zero impact parameter, corresponding to zero orbital angular momentum in the initial state.  We have verified, as intuitively expected, that there are unique (classical) solutions for neutron (and helium) velocities, i.e., there are unique directions for the emitted neutron and helium. Specifically, for a given energy of incident deuteron, there are two solutions that are related by reflection symmetry, see the above discussion.  We have obtained the angular dependence of the emitted neutron and helium as a function of the incident deuterium kinetic energy. Our results may be compared with a quantum calculation~\cite{Kulsrud1982} given by Kulsrud et al. 

\begin{table}
\caption{\label{constantValues}Numerical data taken from Refs.\cite{atomicMasses} and \cite{conversions}}.
\begin{ruledtabular}
\begin{tabular}{lll}
{\rm deuterium mass}  &  $m_d$  & 2.01410177785 {\rm u}    \\
   {\rm tritium mass} &  $m_t$  & 3.01604927767 {\rm u}    \\
{\rm neutron mass}  &  $m_n$  & 1.00866491574 {\rm u}    \\
{\rm ${}^4$He mass}  &  $m_h$  & 4.00260325415 {\rm u}    \\
{\rm speed of light}  &  $c$  & 299792458 {\rm m/s}    \\
{\rm MeV per u}  &   --  & 931.49402823303  {\rm MeV/u} \\
\end{tabular}
\end{ruledtabular}
\end{table}

From our classical relativistic analysis, it seems clear that we could in principle relax the constraint of zero impact parameter, so that we could impose some amount of orbital angular momentum for the deuterium and helium.  Within a classical context (as opposed to a quantum calculation) the amount of orbital angular momentum imposed in the initial conditions is arbitrary and is a continuous quantity (not a discrete quantity, as in quantum calculations)related to the impact parameter.  As described earlier, assuming non-zero angular momentum in the initial state would allow us to treat the case where the deuterium and tritium spins are aligned, which is presumably the more important case. As described earlier, in such as case, we will have orbital angular momentum appear in the products of the collision (neutron and helium).    The  continuum of possible angular momentum in the initial state would then lead to a continuum of angular momentum in the possible final (product) states.  So the neutrons would not have a unqiue direction, but instead, they would have a distribution of possible velocities.  The assumed initial probability distributions of impact parameters would then lead to a probability distributions for neutron velocities and angles.  As described earlier, this is believed to be the more important case.  We leave this for possible future work.

\section{acknowledgments}
This work was sponsored in part by ILIR at the AMRDEC.


\begin{thebibliography}{99}     

\bibitem{Alfassi1990}See for example, Z.B. Alfassi,  {\it Activation Analysis}, Volumes I and II. CRC Press, Boca Raton, Florida, USA, (1990).   See also the web sites, \url{http://archaeometry.missouri.edu/naa_overview.html and http://www.reak.bme.hu/nti/Education/Wigner_Course/WignerManuals/Budapest/NEUTRON_ACTIVATION_ANALYSIS.htm} and \url{http://web.missouri.edu/~umcreactorweb/pages/ac_elemlist.shtml#anchor5}.  

\bibitem{Ohlsen1967}G. G. Ohlsen, Phys. Rev. {\bf 164} 1268 (1967).

\bibitem{Bonar1968}D. C. Bonar, C. W. drake, R. D. Headrick, V. W. Hughes, Phys. Rev. {\bf 174}, 1200 (1968).

\bibitem{Kulsrud1982}R. M. Kulsrud, H. P. Furth, and M. Goldhaber, Phys. Rev. Lett. {\bf 49}, 1248 (1982).

\bibitem{Synge1965}J. L. Synge, {\it Relativity:  The Special Theory}, Second Edition, North-Holland (1965).

\bibitem{atomicMasses}A. H. Wapstra, G. Audi, and C. Thibault. Nuclear Physics {\bf A729}, 129 (2003), see    \url{http://www.nndc.bnl.gov/nsr/nsrlink.jsp?2003WA32,B} and also G. Audi, A.H. Wapstra, and C. Thibault. Nuclear Physics {\bf A72}9, 337 (2003), see \url{http://www.nndc.bnl.gov/nsr/nsrlink.jsp?2003AU03,B}.  Also, see the Los Alamos web page \url{http://www.nndc.bnl.gov/masses/mass.mas03}.

\bibitem{conversions} See for conversion factors, \url{http://physics.nist.gov/cuu/Document/factors_2006.pdf}

\end{thebibliography}
\end{document}